\documentclass[a4paper,twocolumn,amsmath,aps,floatfix]{revtex4}        

\usepackage{graphicx}

\usepackage[margin=10pt,font=small,labelfont=bf,justification=centerlast]{caption}
\usepackage{morefloats}

\begin{document}

\title{Influence of mechanical properties on mixing and segregation of granular material}
\author{Algis D\v{z}iugys}
\email{dziugys@isag.lei.lt}
\author{Robertas Navakas}
\email{rnavakas@mail.lei.lt}
\affiliation{Lithuanian Energy Institute \\ Laboratory of Combustion Processes \\ Breslaujos~g.~3, LT-44403~Kaunas, Lithuania}

\begin{abstract}

Processes of mixing and segregation in a packed bed of granular material stirred by a periodically moving rectangular bar 
are simulated by the discrete element method (DEM). Influence of mechanical properties of the particle material 
(dynamic friction coefficient, density, elastic modulus) on the segregation intensity is researched. Dynamic friction coefficient was found to have the most noticeable influence to the segregation intensity.

\end{abstract}


\maketitle

\newcommand{\vect}[1]{\mathbf{#1}}

\newcommand{\rd}{\mathrm{d}}
\newcommand{\rn}{\mathrm{n}}
\newcommand{\rt}{\mathrm{t}}

\newlength{\fwidth}
\setlength{\fwidth}{\columnwidth}

\section{Introduction}

Granular materials of various forms are ubiquitous in nature and technology. They exhibit a rich variety of phenomena and,
depending on circumstances, can have properties similar to either solids, liquids or gases \cite{Jaeger1996:68:4}, or behave in a completely
different manner.
Size segregation is a characteristic feature of the moving granular media \cite{Kudrolli2004:67}
and it is important,
as well as mixing, in industrial handling of granulated materials. A number of physical mechanisms were suggested to
explain the segregation process \cite{Schroeter2006}. Existing theories of continuous mechanics or statistical physics have only limited applications for
description of granular media, and a unified theory encompassing all the granular phenomena is still missing
\cite{deGennes1998:261,Kadanoff1999:71:1}.
An ability to describe and predict the conduct of granulated materials is important for technological applications in 
industry where bulk materials are routinely handled. On the other hand, granular media is an interesting object for
theoretical studies because of its intricate phenomenology despite ostensibly simple purely mechanical nature.
Numerical simulations can provide a useful insight into the origins of various aspects of this behaviour.

Research data about the influence of the mechanical properties of granular matter to the segregation process is still 
scarce. Besides, simplified models are commonly used: binary mixtures of particles are researched, binary collisions 
between the particles are assumed. Experiments involving the particles with different mechanical properties are difficult
to implement, but numerical simulations with different parameter sets can be performed easily, even though they are
lengthy and require much computing power.

The objective of the present paper is to investigate the mixing process inside granular material and the dependency of mixing
on the properties of the particle material. Most research performed in segregation of granular materials dealt with the situation when the researched material is
put to motion by a moving bottom wall \cite{Kudrolli2004:67}. We modelled a different setup where the material is stirred by a rectangular bar
moving forward and backward in the horizontal direction and buried inside the material, 
which is a model of a simplified case of real equipment, such as forward or backward moving grates widely used in 
industry. Mixing of granular material is sustained by periodic motion of a rectangular
bar inside a rectangular container. We observe the segregation and stratification of particles by their sizes and
intensity of mixing.

\section{Simulation model}

A number of models have been applied for describing granular media with a varying degree of success
\cite{Dziugys2001:3}.
The discrete element method (DEM) proved to be the most accurate, besides, it is simple to implement.
In this method, the motions and collisions of each separate particle are tracked using the equations of
classical mechanics. The equations of motion are integrated with a constant step size considerably shorter than
the duration of the collision. Particle motion obeys the usual Newton equations
\begin{equation}
\label{eq:motion}
\frac{\rd \vect{x}_i}{\rd t}=\vect{v}_i, \: \frac{\rd \vect{v}_i}{\rd t} = \frac{\vect{F}_i}{m_i}, \:
\frac{\rd \vect{w}_i}{\rd t} = \frac{\vect{T}_i}{I_i}, \: i \in [1, N],
\end{equation}
where $\vect{x}_i$ is the $i$-th particle position, $\vect{v}_i$ is its velocity, 
$\vect{F}_i = \sum\limits_{j \neq i} \vect{F}_{ij}+m_i\vect{g}$ is the total force
acting upon the $i$-th particle, $\vect{F}_{ij}$ is the force acting upon
the $i$-th particle arising from its collision with the $j$-th particle,
$m_i$ is the particle mass, $\vect{g}$ is the free-fall acceleration vector, $\vect{w}_i$ is the angular velocity,
$\vect{T}_i = \vect{d}_{cij} \times \vect{F}_i$ is the torque acting upon the particle, 
$\vect{d}_{cij}$ is the collision point vector defined below,
$I_i$ is the moment of inertia which is scalar for spherical particles, $N$ is the total number of particles.
The orientations of the particles are not updated because this has no sense for spherical shape.

\begin{figure}
\includegraphics[width=\fwidth,keepaspectratio]{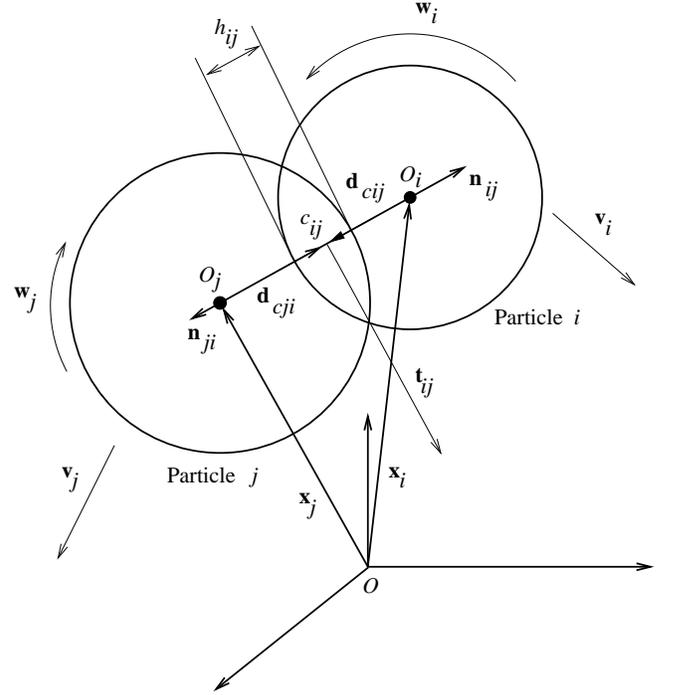}
\caption{Geometrical scheme of the elastic interaction between the colliding particles}
\label{fig-collision}
\end{figure}

The scheme of collision between two particles is shown in Fig.~\ref{fig-collision}.
Particle deformations during collisions are approximated by a partial overlap of their geometric shapes, with the
overlap depth $h_{ij}$. The position of colliding particles $i$ and $j$ in space and their motion is defined by the 
radii-vectors $\vect{x}_i$, $\vect{x}_j$ of their centres  $O_i$, $O_j$, linear velocities $\vect{v}_i$, $\vect{v}_j$ 
and angular velocities $\vect{w}_i$, $\vect{w}_j$. The relative position of the colliding particles is
$\vect{x}_{ij}=\vect{x}_i-\vect{x}_j$. The point of contact $C_{ij}$ is located in the middle of the overlap area and 
its radius-vector is $\vect{x}_{cij}$. The unit vector $\vect{n}_{ij}$ is normal to the contact surface and directed 
toward the particle $i$, $\vect{n}_{ij}=-\vect{n}_{ji}$. 
The vectors $\vect{d}_{cij}$ and $\vect{d}_{cji}$ specify the position of the contact point $C_{ij}$ with respect to 
the centres of the particles $i$ and $j$: $\vect{d}_{cij}=\vect{x}_{cij}-\vect{x}_i$. Relative velocity of the particles
at the contact point is defined as $\vect{v}_{ij}=\vect{v}_{cij}-\vect{v}_{cji}$, $\vect{v}_{ij}=-\vect{v}_{ji}$, 
where $\vect{v}_{cij}=\vect{v}_i+\vect{w}_i \times \vect{d}_{cij}$,
$\vect{v}_{cji}=\vect{v}_j+\vect{w}_j \times \vect{d}_{cji}$ are the velocities of the particles $i$ and $j$ at the 
contact point. Normal and tangential components of the relative velocity are expressed as 
$\vect{v}_{\rn,ij}=(\vect{v}_{ij} \cdot \vect{n}_{ij})\vect{n}_{ij}$, 
$\vect{v}_{\rt, ij}=\vect{v}_{ij}-\vect{v}_{\rn,ij}$, respectively. The contact forces act at the contact point.
The colliding particles can slip with respect to each other in the tangential direction by the distance
$\vect{\delta}_{\rt,ij}=\left|\int \vect{v}_{\rt,ij}(t)\,\rd t\right|$
which is referred to as the integrated slip. As the particles slip, the value of $\vect{\delta}_{\rt,ij}$ increases
until the tangential force exceeds the force caused by static friction, and after that, the tangential force component 
is defined by the dynamic friction. 

There is still no definite agreement what force model best describes the real interaction of the colliding particles 
\cite{Dziugys2001:3,Thomas2004:27:6}. Two most popular models of the normal forces arising between the colliding 
particles $i$ and $j$ are the Hook's law and the Hertz's law. These laws are expressed respectively as
\begin{equation}
\label{eq:Hook-law}
\vect{F}_{\rn,ij}^\mathrm{Hook}=k_{\rn}r_{ij}h_{ij}\vect{n}_{ij}-m_{ij}\gamma_{\rn}\vect{v}_{\rn,ij},
\end{equation}
and
\begin{equation}
\label{eq:Hertz-law}
\vect{F}_{\rn,ij}^\mathrm{Hertz}=k_{\rn}r_{ij}h_{ij} \sqrt{r_{ij}h_{ij}}\vect{n}_{ij}-m_{ij}\gamma_{\rn}\vect{v}_{\rn,ij},
\end{equation}
It can be expected that the effect of a particular force model will be ``averaged out'' over the vast number of collisions
during the process and the differences between these models will not have a significant effect on the statistical parameters
of the granular matter. In order to verify this, we performed the simulations using both models and compared the results
that will be presented below.

The tangential component of the interaction force is expressed as
\begin{equation}
\vect{F}_{\rt,ij}=-\vect{t}_{ij} \min \left[
        \begin{array}{c}
        \mu \left| \vect{F}_{\rn,ij}\right|, \\
        \left|-\gamma_{\rt}m_{ij}\vect{v}_{\rt,ij} -
        k_{\rt} \vect{\delta}_{\rt,ij} \sqrt{r_{ij} h_{ij}}\right|
        \end{array}
        \right],
\end{equation}
where $k_{\rn}=\dfrac{4}{3} \cdot \dfrac{E}{2(1-\nu)}$, $E$ is the elastic modulus of the particle material,
$\nu$ is the Poisson's ratio, $r_{ij}=\dfrac{r_i \cdot r_j}{r_i+r_j}$, $m_{ij}=\dfrac{m_i \cdot m_j}{m_i+m_j}$ are
the normalized radius and normalized mass, $h_{ij}$ is the overlap depth of the particle shapes during collisions 
\cite{Dziugys2001:3}, $\vect{n}_{ij}$ and $\vect{t}_{ij}$ are the normal and tangential vectors at the contact
point, $\gamma_{\rn}$ and $\gamma_{\rt}$ are the normal and tangential (shear) dissipation coefficients,
$\mu$ is the dynamic friction coefficient, $k_{\rt}=\dfrac{8}{3} \cdot \dfrac{G}{2-\nu}$,
$G$ is the shear modulus.
The simulations described below used the time-driven DEM implementation
\cite{Dziugys2001:3,Balevicius2005:34:1}. The equations of motion (\ref{eq:motion}) were integrated using the
6$^\mathrm{th}$-order Gear predictor-corrector scheme \cite{Allen1990}. In order to reduce the amount of resources 
required for collision detection, the simulation area was partitioned into rectangular cells, with the length of the 
cell edge equal to the largest particle diameter. During each time iteration, initially all the particles were assigned
to the corresponding cell according to their current coordinates (Verlet lists). It is evident that only the particles belonging to the
same cell or the nearest neghbouring cells can collide; thereby the number of pairwise collision tests can be reduced
from $O(N^2)$ to almost $O(N)$.

\section{Characterization of mixing and segregation}

The approach to characterization of mixing of particles in granular media stems from the theory of turbulence in fluid 
dynamics \cite{Dziugys2005:60}. It is assumed that the velocity of each particle $\vect{v}_i$ can be split into
its mean value $\vect{V}_i$ and the fluctuating part $\vect{v}'_i$:
\begin{equation}
\label{eq:fluct-velocity}
\vect{v}'_i(t)=\vect{v}_i-\vect{V}_i(t).
\end{equation}
The mean velocity is defined by averaging the velocities of particles surrounding the given $i$-th particle within a
certain preset volume:
\begin{equation}
\vect{V}_i(t)=\frac{\sum\limits_{j=1}^N \vect{v}_j(t) \delta \left(\left|\vect{x}_j-\vect{x}_i\right| \leq R_V\right)}
        {\sum\limits_{j=1}^N \delta \left(\left|\vect{x}_j-\vect{x}_i\right| \leq R_V \right)}.
\end{equation}
Here $R_V$ is the radius of the selected volume surrounding the $i$-th particle over which the averaging is carried out,
$\vect{x}_i$ is the position vector of the $i$-th particle, and a logical function $\delta$ is introduced for brevity:
\begin{equation}
\label{eq:delta-def}
\delta(\mathrm{condition})=\left\{
\begin{array}{ll}
1, & \text{if condition is satisfied} \\
0, & \text{otherwise.}
\end{array}
\right.
\end{equation}
The fluctuating velocity shows how quickly the $i$-th particle changes its position with respect to the surrounding 
particles. By averaging ${\vect{v}'}_{x_n}^2$ and ${\vect{v}'}_i^2$ over the simulation volume, the time-dependent mixing
parameters are introduced:
\begin{subequations}
\label{eq:mixing-PI}
\begin{equation}
\Pi_{x_n}(t)=\sqrt{\frac{1}{N} \sum\limits_{i=1}^N {\vect{v}'}_{x_n,i}^2(t)}, \: n \in [1,2,3],
\end{equation}
\begin{equation}
\Pi(t)=\sqrt{\Pi_x^2(t)+\Pi_y^2(t)+\Pi_z^2(t)},
\end{equation}
\end{subequations}
where the coordinate axes are denoted as $x \equiv x_1$, $y \equiv x_2$, $z \equiv x_3$. Averaging over the
simulation time period $[t_1,t_2]$ yields the mixing parameters during this simulation period:
\begin{subequations}
\label{eq:mixing-PB}
\begin{eqnarray}
P_{B,x_n}=&&\sqrt{\frac{1}{t_2-t_1} \int\limits_{t_1}^{t_2} \frac{1}{N} \sum\limits_{i=1}^{N} {\vect{v}'}_{x_n,i}^2(t)\,\rd t}= \nonumber \\
=&&\sqrt{\frac{1}{t_2-t_1}\int\limits_{t_1}^{t_2} \Pi_{x_n}(t) \, \rd t},
\end{eqnarray}
\begin{equation}
P_B=\sqrt{P_{B,x}^2+P_{B,y}^2+P_{B,z}^2}.
\end{equation}
\end{subequations}

The segregation is estimated quantitatively as the change of vertical positions of the particles during the simulation
period depending on particle sizes. Particles whose radii are within the upper 20\% 
of the total interval of the particle radii are considered ``large'', and those with the radii within the lower 20\% 
of the total interval are considered ``small'', i.e., the interval of radii of small particles is [0.005~m, 0.007~m],
and that of the large particles [0.013~m, 0.015~m]. The average vertical position of particles whose radii are in the 
interval $r_1 \leq r_i \leq r_2$ is defined as
\begin{equation}
\label{eq:y}
\langle y_{r_1,r_2}(t) \rangle = 
        \frac{\sum\limits_{i=1}^N y_i(t) \delta \left(r_1 \leq r_i \leq r_2\right)}
                {\sum\limits_{i=1}^N \delta \left(r_1 \leq r_i \leq r_2\right)}.
\end{equation}
The changes of the average vertical position $\langle y_i \rangle$ of ``small'' and ``large'' particles are compared 
to the change of the average vertical position of all the particles (including large and small ones). However, as the stirring bar moves forward and backwards, the height of the volume occupied by the particles
(``packed bed'') changes. Therefore, it is reasonable to introduce the average height of this volume $H_\mathrm{PB}(t)$ and
to normalize the vertical positions of the particles to this height, in order to discriminate the vertical motion of
individual particles from the motion of the entire packed bed as a whole:
\begin{equation}
\label{eq:norm-y}
\langle Y_{r_1,r_2}(t) \rangle = \frac{\langle y_{r_1,r_2}(t) \rangle}{H_\mathrm{PB}(t)}.
\end{equation}

\section{Simulation results and discussion}

\begin{figure}
\begin{centering}
\includegraphics[width=0.7\fwidth,keepaspectratio]{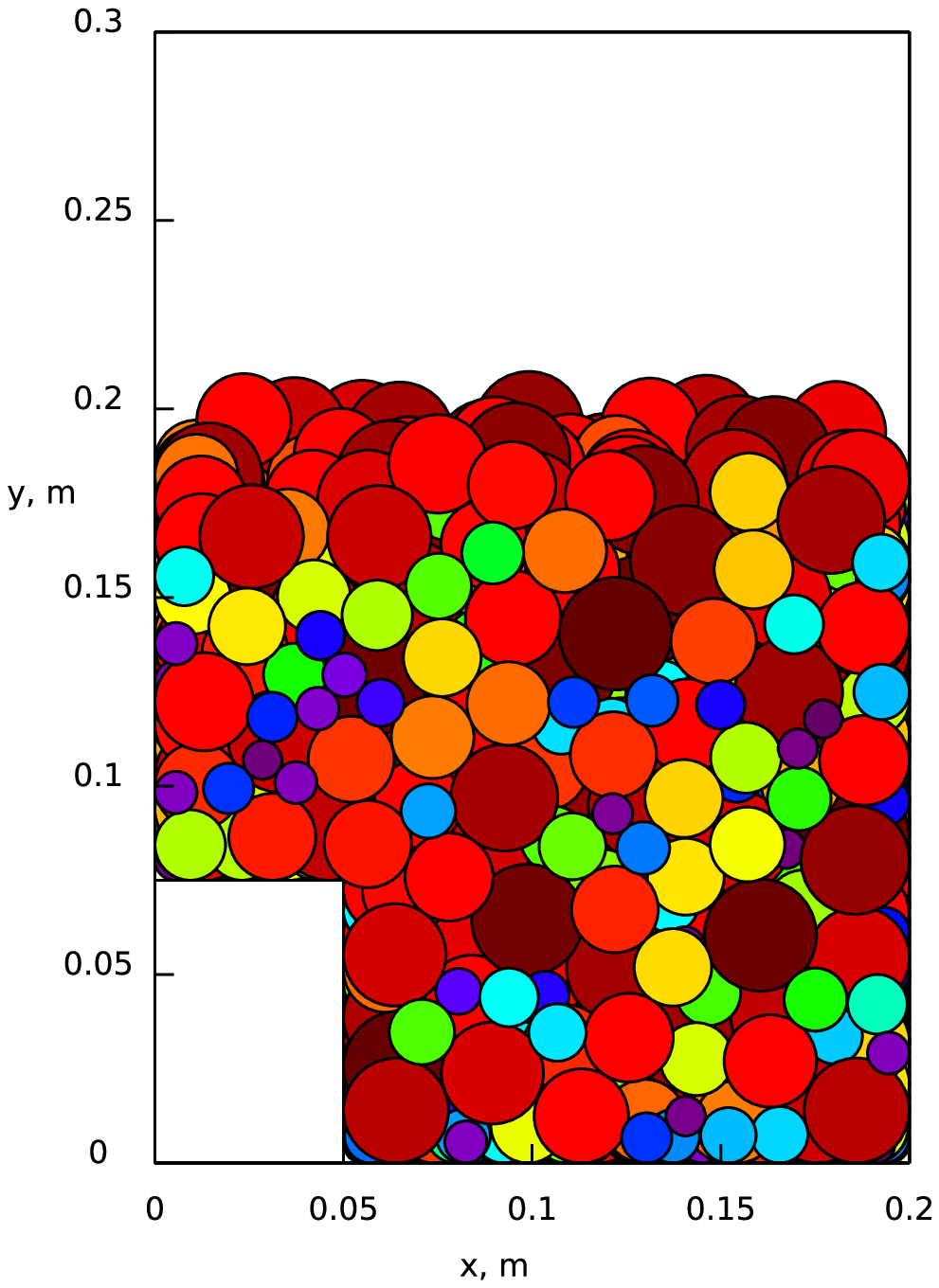}
\includegraphics[width=0.7\fwidth,keepaspectratio]{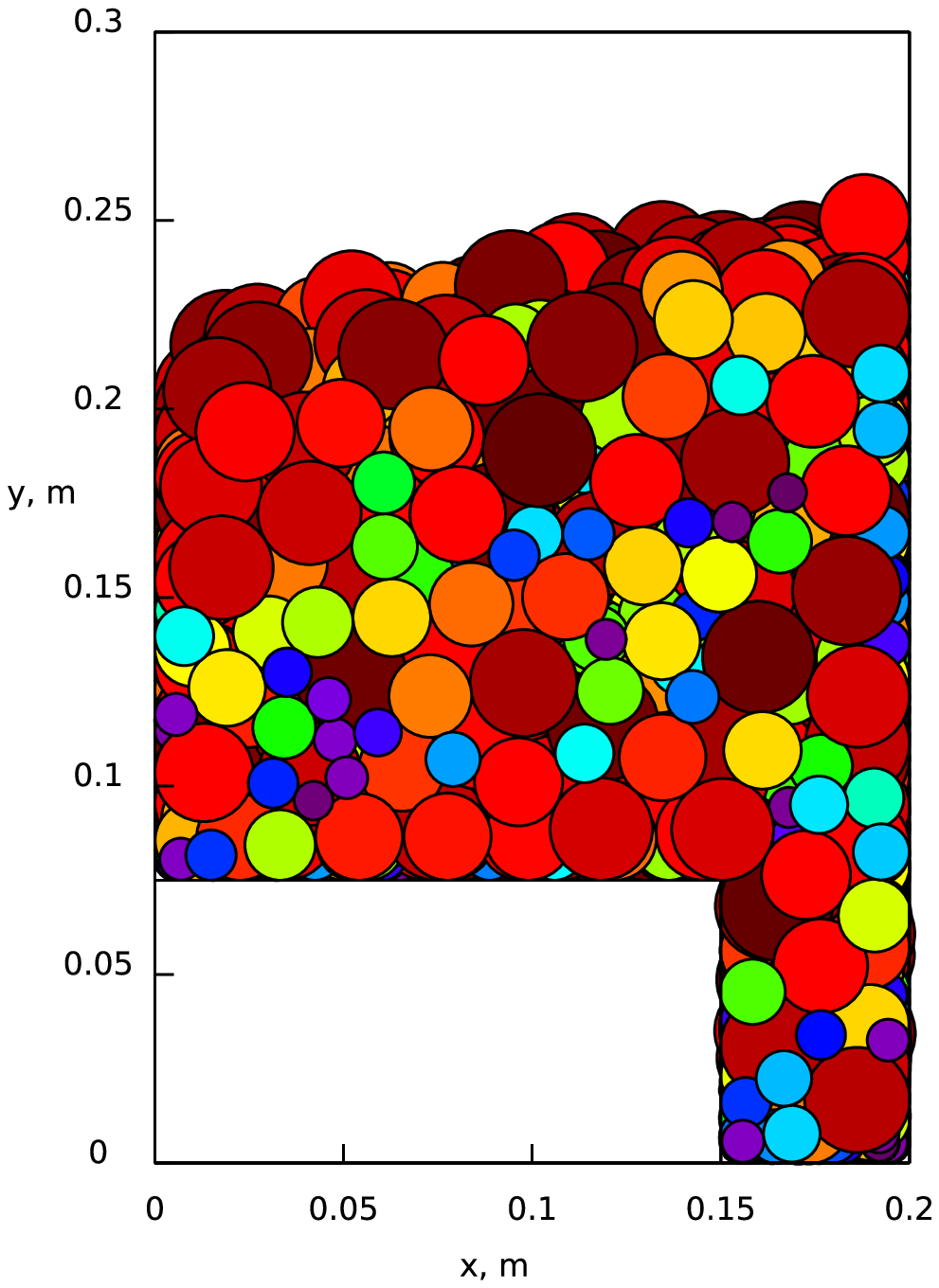}
\includegraphics[width=0.7\fwidth,keepaspectratio]{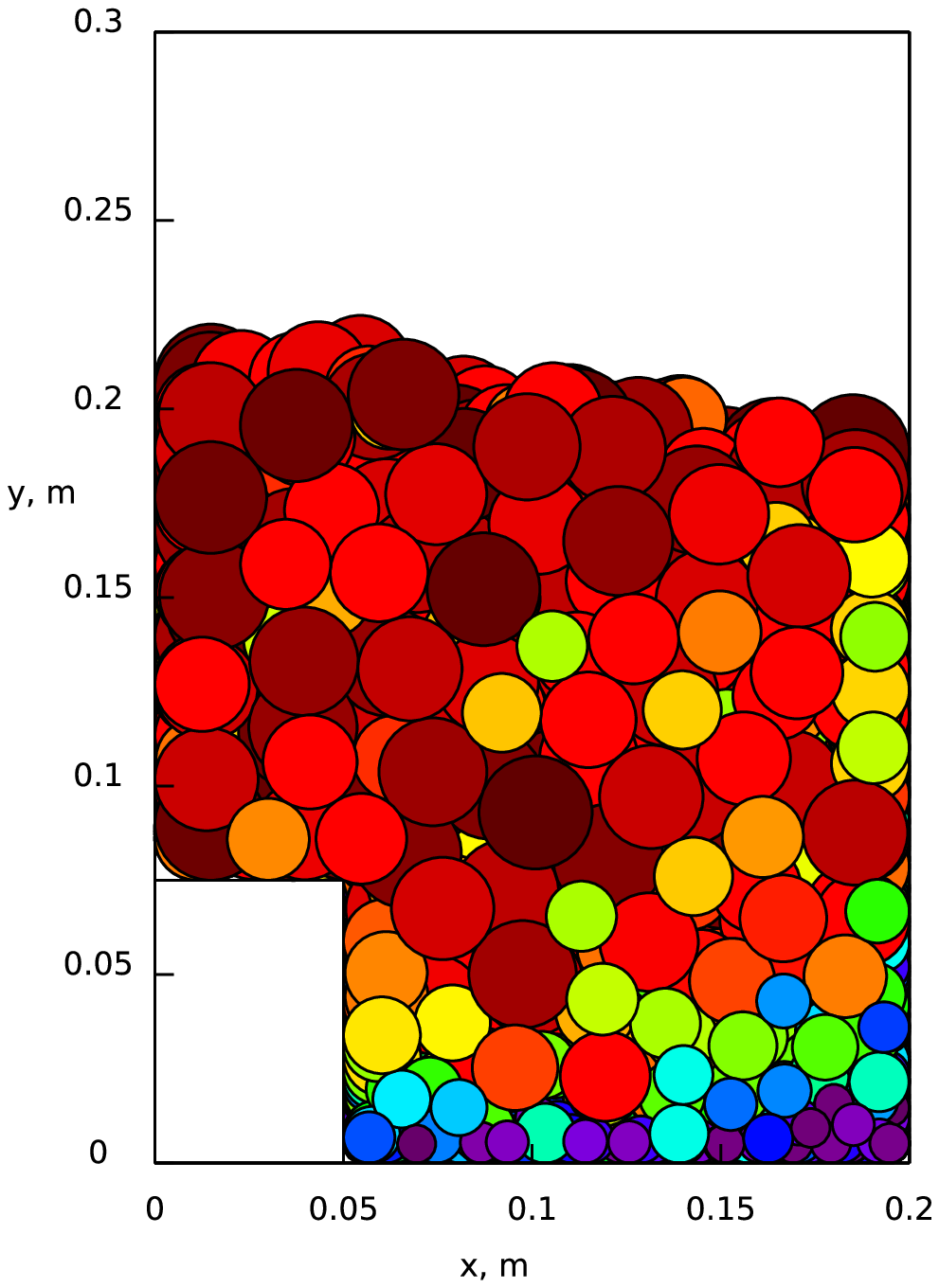}
\caption{\label{fig:positions}
Evolution of particle positions during the mixing process (top to bottom): initial positions at $t=0$~sec,
positions at $t=10$~sec, and at $t=400$~sec. Particles of different diameters are shown in different colors.
}
\end{centering}
\end{figure}
\begin{figure}
\begin{centering}
\includegraphics[angle=-90,width=\fwidth,keepaspectratio]{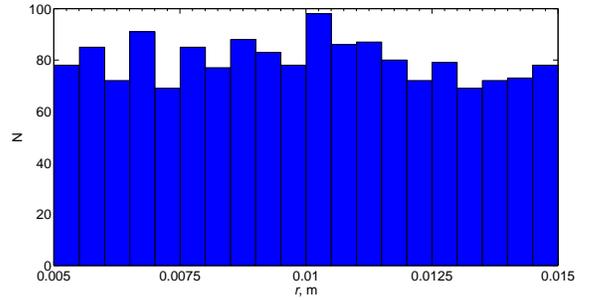}
\caption{\label{fig:size-distribution}
Distribution of particle radii in the simulated system.}
\end{centering}
\end{figure}

The simulated system consists of a rectangular box whose size is $0.2 \times 0.4 \times 0.2$~m$^3$ containing $N=1600$
spherical particles of
various sizes with random distribution of particle depths for particles of different sizes, with radii ranging from 
0.005~m to 0.015~m.  At the 
lower left corner of the box, a rectangular moving bar is located whose width (in the direction of the $z$ axis) is 
equal to the width of the box (Fig.~\ref{fig:positions}). The histogram of particle size distributions is shown in Fig.~\ref{fig:size-distribution}. During the process, the bar moves from its leftmost position to the right along the $x$
axis over the distance of 0.1~m at a constant velocity for the time period $T$, then retracts back to its initial 
position during the same time period $T$, and then the process repeats periodically during the entire simulation time.
The time of motion of the bar from the leftmost to the rightmost position was usually $T=10$~sec, if not specified
otherwise; the time of backward motion is equal to the time of forward motion. The bar motion results in stirring of
the particles and their redistribution in the volume of the box.

The simulations were performed for particles that had initially the same initial positions, linear and angular velocities
for each run. The ``standard'' mechanical parameters of the particle material are listed in Table~{\ref{tbl:prms}};
only the value of the parameter whose influence was to be determined was set different for the corresponding run. The
set of equations (\ref{eq:motion}) was then solved for a certain period of the ``simulated'' time, yielding the positions and velocities of each particle at the
time moments separated by a certain time step. The duration
of the simulation period usually varied from 100~s to 400~s long.
From the resulting data, the mixing and segregation parameters (\ref{eq:mixing-PI}) -- (\ref{eq:norm-y}) 
were calculated.

\begin{table}
\caption{\label{tbl:prms}Standard mechanical properties of the particle material}
\begin{tabular}{|l|c|rl|}
\hline
Parameter & Notation & \multicolumn{2}{c|}{Value} \\
\hline
Minimal particle radius & $r_\mathrm{min}$ & 0.005 & m \\
Maximum particle radius & $r_\mathrm{max}$ & 0.015 & m \\
Density & $\varrho$ & 700 & kg$/$m$^3$ \\
Elastic modulus & $E$ & $10^7$ & Pa \\
Poisson modulus & $\sigma$ & 0.2 & \\
Normal dissipation coefficient & $\gamma_{\rn}$ & 100 & s$^{-1}$ \\
Shear dissipation coefficient & $\gamma_{\rt}$ & 100 & s$^{-1}$ \\
Dynamic friction coefficient & $\mu$ & 0.5 & \\
Shear modulus & $G$ & $3 \times 10^6$ & Pa \\
Period of the stirring bar motion & $T+T$ & 10+10 & s \\
Free fall acceleration module & $\left| \vect{g}\right|$ & 10 & m$/$s$^2$ \\
\hline
\end{tabular}
\end{table}

\begin{figure}
\begin{centering}
\includegraphics[angle=-90,width=\fwidth,keepaspectratio]{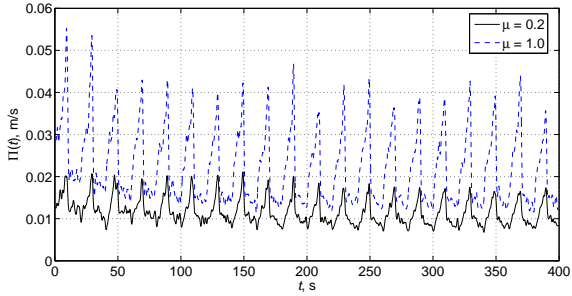}
\caption{
\label{fig:pi-vs-mu}
Evolution of the mixing parameters $\Pi$ for the values of the dynamic friction coefficient
$\mu=0.2$ and $\mu=1.0$.}
\end{centering}
\end{figure}

Segregation of particles can be seen by comparing particle positions at different time moments in Fig~\ref{fig:positions}.
Evidently, smaller particles tend to sink downwards and
accumulate near the bottom as the process progresses. Quantitatively the mixing can be estimated from the evolution 
of the mixing parameters $\Pi(t)$ (Eq.~\ref{eq:mixing-PI}), as shown in Fig.~{\ref{fig:pi-vs-mu}}.

\begin{figure}
\begin{centering}
\includegraphics[angle=-90,width=\fwidth,keepaspectratio]{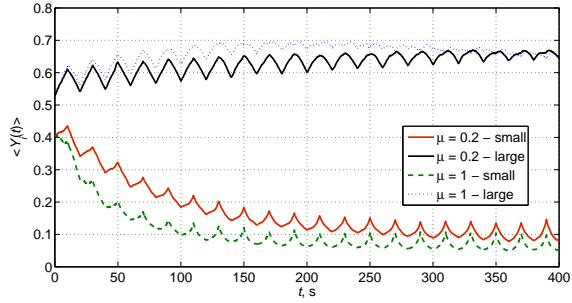}
\caption{\label{fig:y-vs-t-mu}
Evolution of the average vertical positions of small and large particles for the values of the
dynamic friction coefficient $\mu=0.2$ and $\mu=1.0$.}
\end{centering}
\end{figure}

Segregation is demonstrated by the change of the normalized
average vertical positions $\langle Y(t) \rangle$ (Eq.~\ref{eq:norm-y}) of small and large particles
(Fig.~\ref{fig:y-vs-t-mu}). It shows evidently how the small and large particles separate vertically
during the mixing process. The jagged shape of the curve corresponds to upward and downward motion of the
particles following the motion of the bar. As the process progresses, the particles tend to approach a certain nearly 
stationary state where the fluctuations of their positions are influenced only by the motion of the stirring bar. These
fluctuations are more pronounced for small particles accumulating near the bottom, because the available volume changes
as the bar moves. The particles accumulated at the top of the packed bed are located mainly above the bar and are therefore
less influenced by its motion.

The time-averaged mixing coefficients $P_B$, $P_{B,x,y,z}$ (Eq.~\ref{eq:mixing-PB}) are calculated from the time-dependent mixing coefficients
$\Pi(t)$, $\Pi_{x,y,z}(t)$ by averaging over a certain selected time period $\left[t_1, t_2\right]$. As seen from the
temporal evolution of the vertical particle positions, in case of the random initial distribution of particle positions
by their size, mixing is most intense in the beginning of the process, while particles are not yet segregated by their
sizes. Therefore, the values of the parameters $P_B$, $P_{B,x,y,z}$ depend on the selection of the averaging period
$\left[t_1, t_2\right]$. It is possible that some parameters of the material or the mixing procedure have more influence
during the initial stages of the process, while other parameters are more significant as the material is approaching
the steady state.

\begin{figure}
\begin{centering}
\includegraphics[angle=-90,width=\fwidth,keepaspectratio]{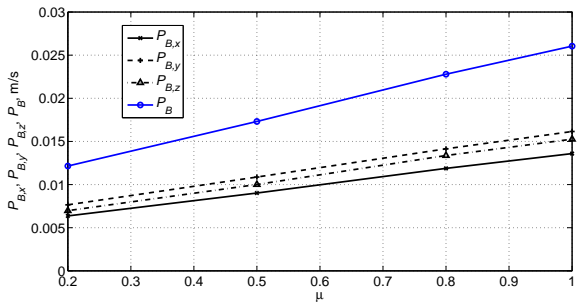}
\caption{\label{fig:PB-vs-mu}
Dependence of the averaged mixing parameters $P_B$ on the dynamic friction coefficient $\mu$.}
\end{centering}
\end{figure}

As seen from Figs.~\ref{fig:pi-vs-mu}--\ref{fig:y-vs-t-mu}, higher values of the dynamic friction coefficient causes more intense mixing. This result is similar to the case of
hydrodynamics where mixing is more intense in more viscous fluids. The dependence of the overal mixing intensity on the
dynamic friction coefficient is nearly linear in the researched range of its values (Fig.~\ref{fig:PB-vs-mu}).

\begin{figure}
\begin{centering}
\includegraphics[angle=-90,width=\fwidth,keepaspectratio]{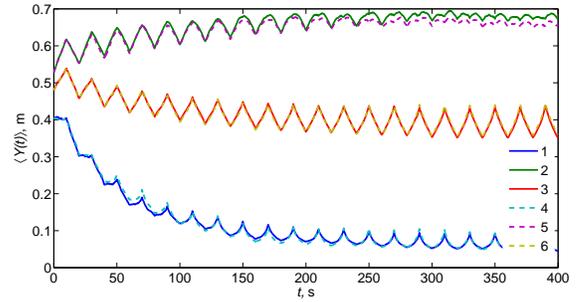}
\caption{\label{fig:y-vs-t-rho}
Evolution of vertical positions of small (curves 1, 4), large (curves 2, 5) and all (curves 3, 6) particles with the
densities of $\varrho=500$~kg$/$m$^3$ (curves 1, 2, 3) and $\varrho=900$~kg$/$m$^3$ (curves 4, 5, 6).}
\end{centering}
\end{figure}

In order to identify the influence of the material density of the particle material, simulations were performed for particles
with the densities $\varrho=500$~kg$/$m$^3$ and $\varrho=900$~kg$/$m$^3$. The changes of the vertical particle positions for
these different values of the density is shown in Fig.~\ref{fig:y-vs-t-rho}. Evidently, the overal effect of the material
density on the segregation process is insubstantial. However, it can be noticed that at the later stages of the process
the larger particles with lower density accumulate slightly higher in the packed bed than in the case of denser particles.
This results in looser packing of the particles in the top layers of the bed.

\begin{figure}
\begin{centering}
\includegraphics[width=\fwidth,keepaspectratio]{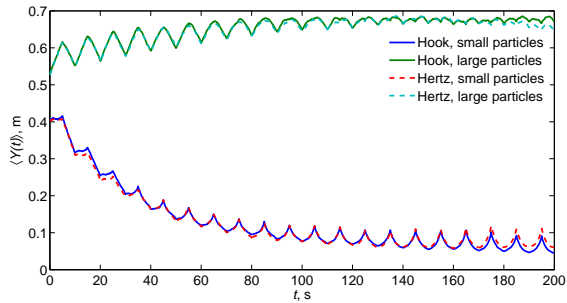}
\caption{\label{fig:Hertz-vs-Hook}
Evolution of normalized average vertical positions $\langle Y(t) \rangle$ of small and large particles interacting
according to the Hook's law and the Hertz's law.}
\end{centering}
\end{figure}

Fig.~\ref{fig:Hertz-vs-Hook} demonstrates the evolution of the vertical positions of particles interacting according to
the Hook's law (Eq.~\ref{eq:Hook-law}) compared to the case of interaction according to the Hertz law
(Eq.~\ref{eq:Hertz-law}). As noted above, the difference between the two force models does not manifest itself 
significantly in the mixing process. The simulations using the Hertz normal force model were performed with the values
of the dynamic friction coefficient $\mu=0.5$ and $\mu=0.8$. Table~\ref{tbl:P-Hertz-vs-Hook} lists the values of the
mixing coefficient $P_B$ obtained for these different values of the dynamic friction coefficient for both Hook and Hertz
models. The values of $P_B$ were calculated by averaging $\Pi(t)$ over the time interval $[1,100]$~s, when the mixing
process is still more intense than in the later stages. The difference between the Hook and Hertz models yields the 
difference in the mixing coefficient of approximately 10\%.
It can be noticed that the difference of this coefficient for different values of the dynamic friction coefficient
is larger in case of the Hertz model; however, more simulations with different parameter sets are required in order to
determine the sensitivity of the Hertz model to the mechanical parameters.

\begin{table}
\caption{\label{tbl:P-Hertz-vs-Hook}
Comparison of the values of the mixing coefficients $P_{B}$ (in cm$/$s) averaged over the period from $t_1=0$~s
to $t_2=100$~s for the cases of interaction according to
the Hook's law and the Hertz's law for different values of the dynamic friction coefficient: $\mu=0.5$ and $\mu=0.8$.}
\begin{tabular}{|l|c|c|}
\hline
Normal force model & $\mu=0.5$ & $\mu=0.8$ \\
\hline
Hook & 1.3569 &  1.6780 \\
\hline
Hertz & 1.2947 & 1.7335 \\
\hline
\end{tabular}
\end{table}

Dependence of the mixing process on the elastic modulus was also investigated in the range
$E=(0.6 \div 1.4) \times 10^7$~Pa. The values of the mixing parameters $P_B$, $P_{B,x,y,z}$ decrease slightly as the values
of $E$ increase, however, this change was insignificant, within 2\%.
On the other hand, the range of the values of $E$ was quite small and the effects of this parameter were probably 
masked by other influences. Simulations in a wider range are necessary to identify the possible influence of this parameter.
The same considerations apply to the Poisson module $\nu$ that enters the equations (\ref{eq:Hook-law}) and
(\ref{eq:Hertz-law}) through the coefficient $k_{\rn}$ together with the elastic modulus $E$.



\section{Conclusions}

The mixing parameters introduced by the hydrodynamic analogy were applied to characterize the mixing and segregation of
granular matter. Segregation of particles of different sizes was estimated by the change of their normalized depths.

The mixing intensity weakly depends on the elastic coefficient of the particle material in the researched range 
of the values. In order to better understand the influence of this parameter, it is necessary to research a wider range 
of its values. On the other hand, the mixing intensity depends noticeably on the value of the dynamic friction 
coefficient. In turn, more intense mixing results in faster segregation. This confirms the validity of the mixing
parameters for estimating the segregation. The dependence of the mixing intensity on the dynamic friction coefficient
is linear in the researched range of the values.

There is a certain difference of the obtained results depending on whether the Hook's or Hertz's model of the contact
force between the colliding particles is used; however, this difference is not substantial. Further research is needed 
to identify possible influence of some parameters, by extending the range of the researched values.

\subsection*{Acknowledgement}

This work has been performed under the project HPC-EUROPA (RII3-CT-2003-506079), with the support of the European 
Community -- Research Infrastructure under the FP6 ``Structuring the European Research Area'' Programme.

\end{document}